\documentclass[pre,twocolumn,showpacs,cite]{revtex4-1}
\usepackage{amsmath,amssymb,amsfonts,graphicx,hyperref}

\begin{document}

\title{Analytical results for long time behavior in anomalous diffusion}

\author{R. M. S. Ferreira$^{1}$\email{rogelma@fisica.ufc.br}, M. V. S. Santos$^{2}$, C. C. Donato$^{2}$, J. S. Andrade Jr.$^{1}$, F. A. Oliveira$^{2}$}.

\affiliation{ $^{1}$Departamento de F\'{\i}sica,Universidade Federal do Cear\'a, Caixa Postal 6030, 60455-900 Fortaleza, Cear\'a, Brazil}
\affiliation{$^{2}$ Instituto de F\'{\i}sica e Centro Internacional de F\'{\i}sica da Mat\'{e}ria Condensada,
Universidade de Bras\'{\i}lia, Caixa Postal 04513, 70919-970 Bras\'{\i}lia, Distrito Federal, Brazil }

\begin{abstract}
We investigate through a Generalized Langevin formalism the phenomenon of anomalous diffusion for asymptotic times, and we generalized the concept of the diffusion exponent. A method is proposed to obtain the diffusion coefficient analytically through the introduction of a time scaling factor $\lambda$. We obtain as well an exact expression for $\lambda$  for all kinds of diffusion. Moreover, we show that $\lambda$ is a universal parameter determined by the diffusion exponent. 
The results are then compared with numerical calculations and very good agreement is observed. The method is general and may be applied to many types of stochastic problem. 
\end{abstract}



\maketitle

\section{\label{sec1} Introduction.}

The study of systems with long range memory reveals some physical phenomena that are still not well understood, especially in systems which are outside the state of equilibrium
or those in which the existence of anomalous diffusion is verified~\cite{Morgado02,Metzler00,Zapperi01,Filoche08,Medino12,Lapas08,Dybiec12,Siegle10}. Here we show a simple analytical  method which describes the behavior of the diffusion for large and intermediate times. In order to do that, we first generalize the concept of the diffusion exponent. Then we present a conjecture to obtain, through the introduction of a time scaling factor $\lambda$, an analytical asymptotic result for the diffusion coefficient for long times. We obtain the 
scaling factor exactly and we show as well its universal behavior.  We derive a numerical method to obtain the correlation function of velocities for an ensemble of particles from any given memory. We compare both methods and we obtain excellent agreement.  The method has general application in the study of stochastic processes and it could be applied to several situations of physical interest.

\section{\label{sec2} Generalized Langevin Equation and Diffusion.}

The generalized Langevin equation (GLE) is a stochastic differential equation which can be used to model systems driven by colored random forces. 
For the velocity operator $v(t)$ this equation can be written  as,
\begin{equation}
m\frac{dv(t)}{dt} = - m\int_0^t \Gamma(t - t')v(t')dt' + \xi(t), \label{eq1}
\end{equation}
where $\Gamma(t)$ is the retarded friction kernel of the system, or the memory function. Here, $\xi(t)$ is a stochastic noise subject to the conditions $\langle\xi(t)\rangle = 0$, $\langle\xi(t)v(0)\rangle = 0$, and
\begin{equation}
C_{\xi}(t) = \langle\xi(t)\xi(0)\rangle = m^2\langle v^{2}(t)\rangle \Gamma(t), \label{eq2}
\end{equation}
where $C_{\xi}(t)$ is the correlation function for $\xi(t)$, and the angular brackets denote an average over the ensemble of
particles. Equation~(\ref{eq2}) is Kubo's fluctuation dissipation theorem (FDT)~\cite{Kubo66,Kubo91}. The presence of the kernel $\Gamma(t)$ allows us to study a large number of correlated processes. In the real world, the vast
majority of problems are non-Markovian, i.e., there is correlation between the various stages of dynamic evolution. This property is what we call memory, and it makes remote events of the past important to
dynamic events in the present time.

 Using the GLE it is possible to study the asymptotic behavior of the second moment of the particle movement,
\begin{equation}
\lim_{t\rightarrow\infty} \langle x^{2}(t)\rangle =2 D(t)t \sim {t^\alpha}, \label{eq4}
\end{equation}$  $
 which characterizes the type of diffusion presented by the system. Here, $D(t)$ is the diffusion coefficient as a function of time.
 Moreover,  for   an asymptotic behavior of the form
\begin{equation}
\lim_{t\rightarrow\infty}  \langle x^{2}(t)\rangle  \sim {t^{\alpha}}{ [ \ln{(t)}] ^{ \pm 1}}, \label{x2g}
\end{equation}
we shall define respectively an 
$ \alpha^{\pm}$  diffusive behavior. Here  the exponent $ \alpha= \alpha^{\pm}$  arises in analogy with the critical exponents in a phase transition. For example, in the two-dimensional Ising model the critical exponent for the specific heat  is $ \alpha= 0^{+}$ because it does not have  a power law behavior; rather it has $\ln|T-T_c|$ behavior for temperatures $T$ close to the transition temperature $T_c$.  This generalized nomenclature is pertinent here  since there are quite a large number of possibilities of combinations for logarithmic and power law behaviors.

 In this way the behavior of $D(t)$ can be determined using
 \begin{equation}
\lim_{t\rightarrow\infty} D(t) = \lim_{t\rightarrow\infty} \lim_{z\rightarrow 0} \int_0^{t} C_v(t')\exp(-z t')dt'= \lim_{z\rightarrow 0}\widetilde{R}(z), \label{eq9}
\end{equation}
where $R(t)=C_v(t)/C_v(0)$, with $C_v(0)=1$,  and $\widetilde{R}(z)$ is the Laplace transform of $R(t)$. For $t\rightarrow\infty$ and normal diffusion this is the Kubo formula \cite{Kubo91}. The limits can be justified using the final value theorem (FVT) for a Laplace transform \cite{Gluskin03}, i.e., for any function $g(t)$ with Laplace transform $\widetilde{g}(z)$ then 
$ \lim_{t\rightarrow\infty} g(t) =  \lim_{z\rightarrow 0} z \widetilde{g}(z)$. Now a Laplace transform of the integral gives
$ \widetilde{D}(z) = \widetilde{R}(z)/z$, and we end up with the equation above.

 Now we multiply  Eq.~(\ref{eq1}) by $v(0)$ and  take the average over the ensemble, with $\langle\xi(t)v(0)\rangle = 0$, to obtain a self-consistent equation for $R(t)$ in the form
\begin{equation}
\dot{R}(t) = -\int_0^t \Gamma(t - t')R(t')dt' \label{eq6}.
\end{equation}
We then  Laplace transform  Eq.~(\ref{eq6}) to get
\begin{equation}
\widetilde{R}(z)=\frac{1}{z + \widetilde{\Gamma}(z)}. \label{eq7}
\end{equation}

Time correlation functions play a central role in non-equilibrium statistical mechanics in many areas, such as the dynamics of polymeric chains~\cite{Toussaint04}, metallic liquids~\cite{Rahman62}, Lennard-Jones liquids~\cite{Yulmetyev03}, 
 ratchet devices~\cite{Bao03, Bao06},  spin waves \cite{Vainstein05}, Heisenberg ferromagnets and dense fluids \cite{Balucani03}.
Consequently to invert this transform, or a similar one,  is crucial. 
 Unfortunately, in most cases it is not an easy task. In those
situations, the use of numerical methods is an alternative to
overcome this problem.  Our main objective here is to show a process to  obtain the asymptotic behavior analytically.  Although the method can be applied to  several situations, we concentrate here on the analysis of diffusion.

\section{\label{sec3} The conjecture}

 We claim that   after a  ``transient time'' $\tau$, i.e., for  $t> \tau$, the leading term for $D(t)$ will fulfill Eq.~(\ref{eq9}) within a given approximation.  In this context $t\rightarrow\infty$  is equivalent to $t \gg \tau$. Now  we imposed the scaling
\begin{equation}
z \rightarrow \lambda/t . 
\label{zt}
\end{equation}
 In order to determine $\lambda$ we rewrite Eq.~(\ref{eq9}) as
 \begin{equation}
 \lim_{t\rightarrow\infty} D(t)= \lim_{t\rightarrow\infty}\widetilde{R}(z=\lambda/t)=\lim_{t\rightarrow\infty} \frac{t}{f(t)},
  \label{eq9b}
\end{equation}
 where
\begin{equation}
  f(t)= \lambda + t \widetilde{\Gamma}(\lambda/t). \label{ft}
\end{equation}
  The derivative of Eq. (\ref{eq9b}) yields
 \begin{equation}
 \lim_{t\rightarrow \infty}R_1(t)= \lim_{t\rightarrow\infty} \frac{d}{dt} D(t)= \lim_{t\rightarrow\infty} [ 1-t\frac{d}{dt} \ln{[f(t)]} ]/f(t) ,
  \label{R1}
\end{equation}
while from the FVT we have
 \begin{equation}
 \lim_{t\rightarrow\infty}R_2(t)= \lim_{z \rightarrow 0}z \widetilde{R}(z) = \lim_{t\rightarrow \infty}\frac{\lambda}{f(t)}.
  \label{R2}
\end{equation}
The relative diference 
 \begin{equation}
\Delta R(t)  = \frac{R_2 - R_1}{R_2} = [\lambda - 1 + t\frac{d}{dt} \ln{[f(t)]}]/\lambda 
  \label{dR}
\end{equation} 
 should evolve to zero as $t \rightarrow \infty$. For $ \lambda \neq 0$ this yields the exact value 
  \begin{equation}
 \lambda =  1 - \lim_{t\rightarrow\infty}t\frac{d}{dt} \ln{[f(t)]} .
  \label{lambda}
  \end{equation} 
 The scaling works as long as the GLE, Eq. (\ref{eq7}), works. To obtain $ \lambda$   we need more information about $\widetilde{\Gamma}(z)$, which may be different for every system. However, since our interest is in the asymptotic behavior, we can expand $\widetilde{\Gamma}(z)$ , in Taylor or Laurent series  around $z=0$, in the form
\begin{equation}
\widetilde{\Gamma}(z) \sim z^{\nu}[ a -b\ln(z)-c/\ln(z)], \label{Gammag}
\end{equation}
where $a$, $b$, and $c$ are positive constants. Note that we give  especial attention to $\ln(z)$, since it will give us the behavior pointed out in Eq. (\ref{x2g}). For $b=0$ this gives a diffusion with exponent $\alpha$; for $b\neq 0$ this gives an $\alpha ^{-}$, and for $a=b=0$ and $c \neq 0$ we get an $\alpha ^{+}$ diffusion. If $\widetilde{\Gamma}(z)$ has another contribution, besides $\ln(z)$, that can not be expanded at the origin we keep it and expand the other parts. However, most of the memories in the literature can be cast in the form Eq. (\ref{Gammag}) for small $z$. Now we introduce Eq. (\ref{Gammag}) into Eq. (\ref{lambda}) to obtain $\lambda = \nu$ for $\nu <1$, and $\lambda = 1$ for $\nu \geq 1$. Notice that it does not depend on $a$,  $b$, or $c$, which suggests a universal behavior.

 In our conjecture some points deserve attention: First, we are considering integrals, of the form Eq. (\ref{eq9}),
where the function $R(t)$ is well behaved, and limited to
 $ - 1 < R(t) < 1$, since $C_v(t) \leq C_v(0)$. $R(t)$ is such  that it always has a well-defined behavior for finite $t$, even when the integral diverges as $t \rightarrow \infty $, as in superdifusion.
 Second, $D(t)$ must have a leading term as $t \rightarrow \infty $, which determines the diffusion.
For example, the inverse Laplace transform of $\widetilde{R}(z)$ is
\begin{equation}
 R(t)= \frac{1}{2 \pi i} \int_{-i \infty+ \eta}^{+i \infty +\eta} \widetilde{R}(z)\exp(zt)dz.
\end{equation}
Here the real number  $\eta$ is such  that all the singularities lie at the left of the line joining the limits.
Consider now Eq. (\ref{Gammag}) with $b=c=0$, and $\nu \leq1$; then $\lim _{z \rightarrow 0} \widetilde{R} (z) \sim z ^{ -\nu}$, and
\begin{equation}
\lim_{t \rightarrow \infty} R(t) \propto t^{\nu-1}\int_{-i \infty+ \eta'}^{+i \infty +\eta'} s^{-\nu} \exp(s)ds \propto t^{\nu-1}, 
\end{equation}
where we have done the transformations $s = zt$ and $\eta' = \eta/t$. For $\nu>0$ the only pole is at $s=0$ , and the condition in $\eta'$ will be automatically satisfied.  Now by direct integration on Eq. (\ref{eq9})  we obtain $D(t) \propto t^{\nu}$ . From the scaling we get  the equivalent result
\begin{equation}
\lim_{t \rightarrow \infty} D(t) = \lim_{z \rightarrow 0}\widetilde{R}(z=\lambda/t) \sim \lim _{t \rightarrow \infty} \widetilde{R}(\lambda/t) \sim t^{\nu} .
\end{equation}
Note that the above exact result is not only for power laws, but for any function behaving as a power law for large $t$. We confirm as well the relation $\alpha=\nu+1$, obtained by Morgado \textit{et al.} \cite{Morgado02}. Our results can be readily expressed as
\begin{equation}
            \lambda = \alpha -1= \alpha^{\pm}-1=\left\{
                       \begin{array}{ll}
                        \nu ,  \qquad  -1< \nu < 1,\\
                        1,  \qquad \nu \geq  1.\\
                       \end{array} \right . \label{lam}
\end{equation}
The  factor $\lambda$ depends  only on the diffusion exponent $\alpha$, consequently it is universal. Moreover it will be the same for $\alpha$ or $ \alpha^{\pm}$. For normal diffusion $\alpha=1$,  or for $\alpha=1^{\pm}$, $\lambda=0$. However, we still can obtain the final value. Consider as example the Langevin equation without memory; for that we have $R(t)=\exp{(- \gamma t)}$ and $ \widetilde{R}(z)=(\gamma +z)^{-1}$. From Eq. (\ref{eq9b}) we get
\begin{equation}
 \lim_{t \rightarrow\ \infty}D(t)= \lim_{t \rightarrow\ \infty} \widetilde{R}(\lambda/t)=\frac{t}{\gamma t +\lambda}= \gamma^{-1},
\end{equation}
while direct integration gives
\begin{equation}
 \lim_{t \rightarrow\ \infty}D(t)= \lim_{t\rightarrow\infty}  \int_0^{t} R(t')dt'= \gamma^{-1}.
 \end{equation}
 In this case the scaling
 yields correctly the wanted final value.

 
 Equation (\ref{eq6}) imposes as well some requirements on $R(t)$.  First its derivative  must be null at the origin, i.e., the integral in the right hand side must be null at $t=0$. This is true except for
non analytical memories, such as $\delta$ functions. Indeed, we do not expect exponential behavior of the form $R(t)=\exp{(-\gamma |t|)} $ with a discontinuous derivative at the origin \cite{Vainstein06, Lee83} .
 Second, in Eq. (\ref{eq1}), for a bath of harmonic oscillators  the noise can be obtained as \cite{Vainstein06}
\begin{equation}
 \xi(t) = \int \sqrt{2k_BT g(\omega)} \cos [\omega t + \phi(\omega)] d \omega, \label{noise}
\end{equation}
where $ 0 < \phi(\omega) < 2 \pi $ are random phases and $g(\omega)$ is the noise spectral density.
The FDT yields
\begin{equation}
\Gamma(t)=  \int g( \omega) \cos(\omega t) d\omega. \label{Gamman}
\end{equation}
This shows that the memory is an even function of $t$. An analytical extension of $\widetilde{\Gamma}(z)$ in the whole complex plane has the property $\widetilde{\Gamma}(-z)=-\widetilde{\Gamma}(z)$. Consequently, from Eq. (\ref{eq7}), $\widetilde{R}(-z)=-\widetilde{R}(z)$, or $R(-t)= R(t)$.  In short, it requires  well-behaved functions and derivatives. Even functions have zero derivatives at the origin as required before.

\section{\label{sec4} The ballistic diffusion}

Let us  consider the spectral density
\begin{equation}
            g(\omega) = \left\{
                       \begin{array}{ll}
                        b\omega_s^{1-\beta}  \omega^{\beta},  \qquad  \omega \leq \omega_s ,\\
                        0, \qquad \omega > \omega_s .\\
                       \end{array} \right . \label{dens}
\end{equation}

  This is a generalization of the  Debye density of states. Here $b>0$ is a dimensionless constant, and $\omega_s$ is a cutoff frequency.
For $\beta \neq 0$ we get anomalous diffusion. In particular, for $\beta=1$ we introduce Eq. (\ref{dens}) into Eq. (\ref{Gamman}) to obtain
\begin{equation}
 \Gamma(t)= b \omega_s^2  \left(\frac{\sin(\omega_s t)} {\omega_st} +\frac{\cos(\omega_s t)-1}{(\omega_s t)^2} \right), \label{Gammaf}
\end{equation}
with the Laplace transform
\begin{equation}
\widetilde{\Gamma}(z)= \frac{bz}{2} \ln{ \left[1+ \left( \frac{\omega_s}{ z}\right)^2  \right]} .
\label{Gamzf}
\end{equation}
First, we have the analytical function $D(t)=\widetilde{R}(z=\lambda/t)$; second, from Eq. (\ref{lambda}) we obtain  $\lim_{t\rightarrow \infty} \lambda =1$, exactly. This is ballistic diffusion of the form $\alpha=2^{-}$.

\begin{figure}
\resizebox{9cm}{!}{\includegraphics{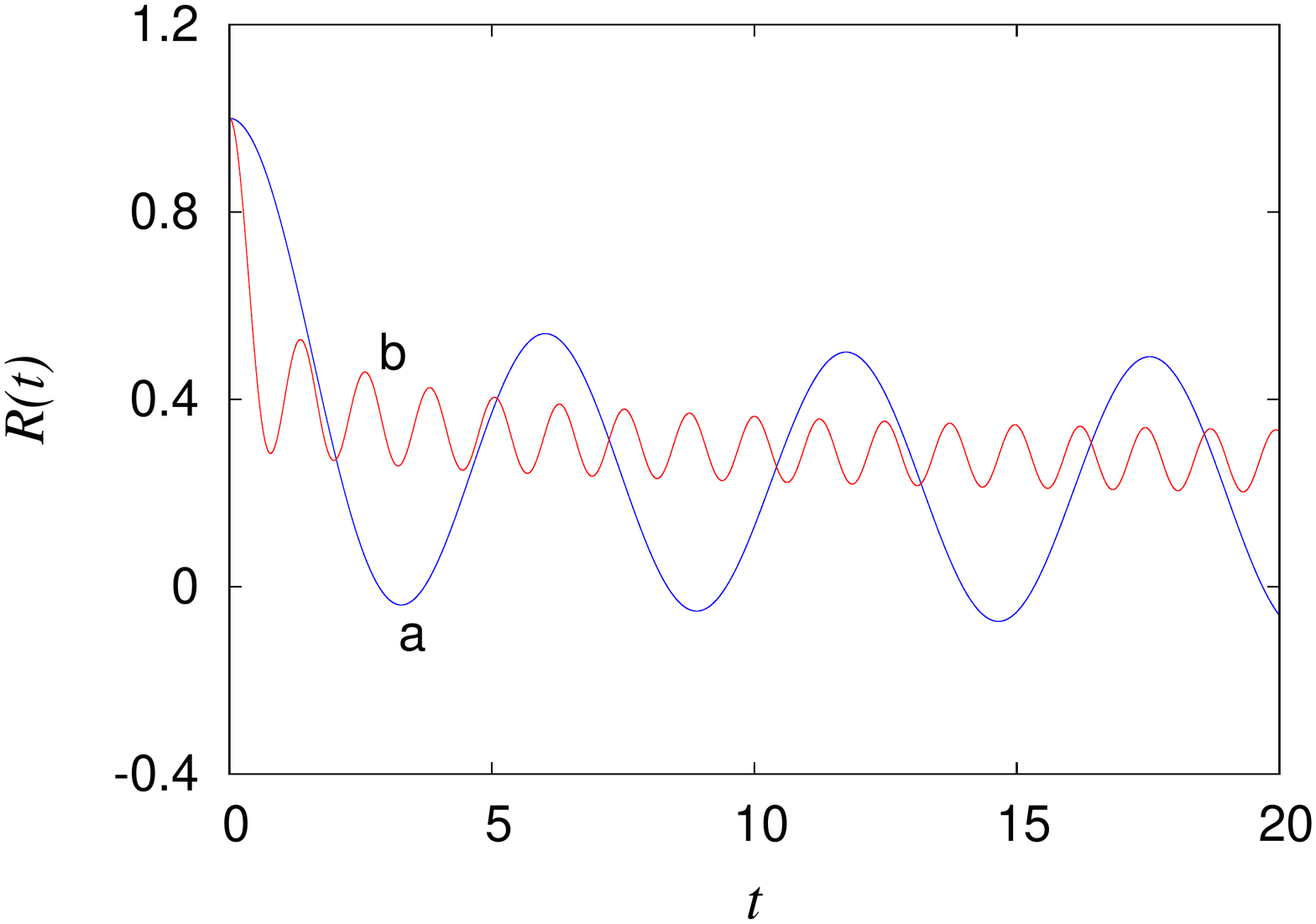}}
\caption{\label{fig1} Correlation function $R(t)$ as a function of time $t$.  We use the memory (\ref{Gammaf}) and numerical integration. Curve \textit{a}, $\omega_s=1$, and $b=1$; curve \textit{b}, $\omega_s=5$, and $b=1/2$.}
\end{figure}

\begin{figure}[t!]
\resizebox{9cm}{!}{\includegraphics{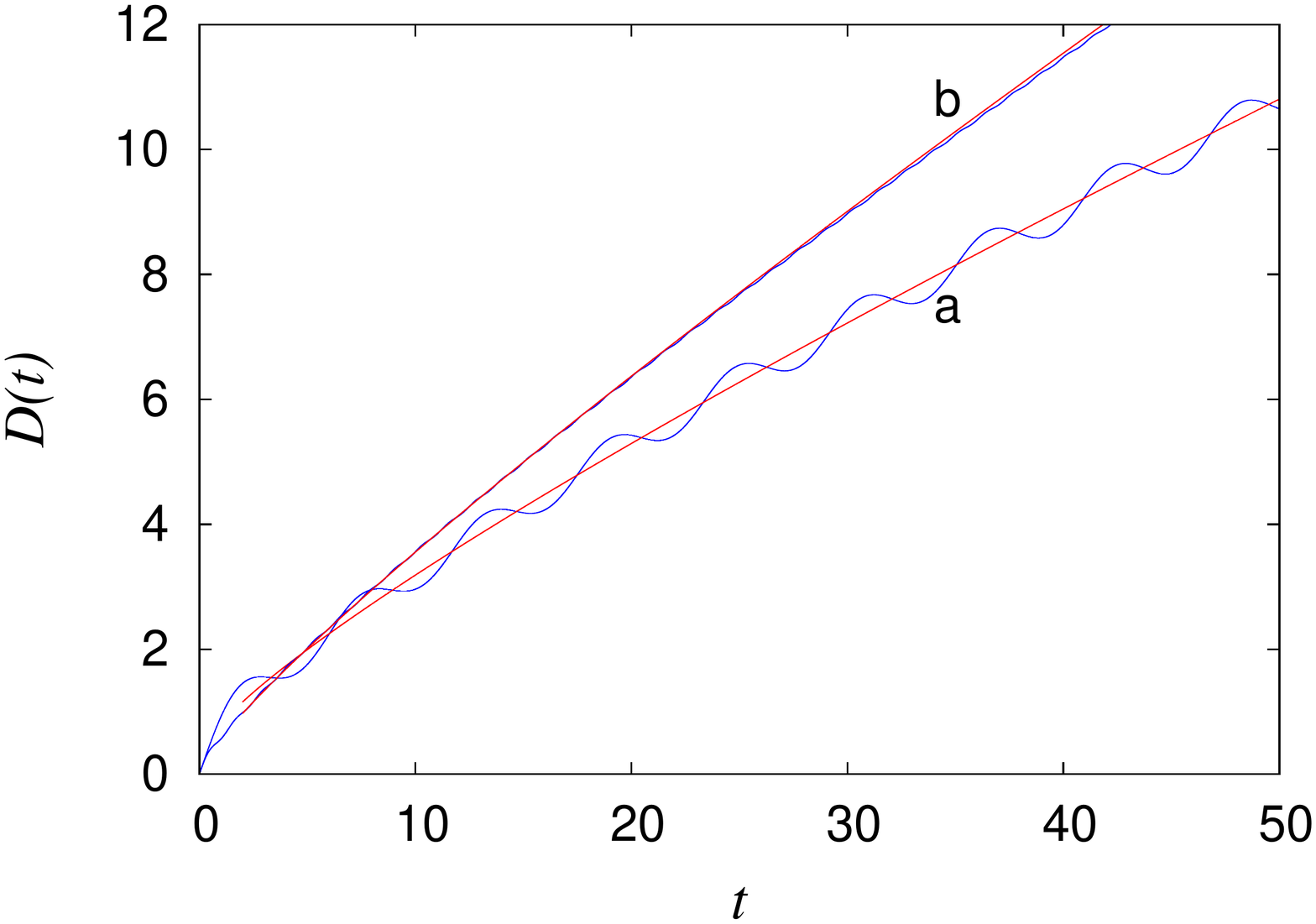}}
\caption{\label{fig2} Diffusion coefficient $D(t)$ as a function of time $t$.  Curve \textit{a}, $\omega_s=1$, and $b=1$; curve \textit{b}, $\omega_s=5$, and $b=1/2$. The oscillatory curves are the numerical result. The  curves without oscillations are the analytical asymptotic limit. We see in curve \textit{b} that the two curves collapse onto a single one. }
\end{figure}
    Now we compare the analytical asymptotic with a numerical solution of Eq.~(\ref{eq6}).  To do this, we rewrite this equation in a discrete form, and then we
expand it up to terms of order $\Delta t^{2n}$ to obtain
\begin{equation}
R(t+\Delta t) = R(t-\Delta t) + 2\sum_{k=0}^n R^{(2k-1)}(t) \frac{ (\Delta t)^{2k-1}}{(2k-1)!}  \label{eq14},
\end{equation}
where $R^{(n)}(t)$, is the time derivative of $R(t)$ of order $n$.
Note that this expansion eliminates all the even derivatives. Now we can obtain all $R(t+\Delta t)$ from the sequence of the previous value of $R(t)$, starting from $R(0)=1$. From these values, its possible to get the diffusion coefficient through direct integration of Eq.~(\ref{eq9}).

In Fig.~\ref{fig1} we plot the correlation function $R(t)$ as a function of time $t$. The curves correspond to the numerical solution, and  are calculated
using  Eq.~(\ref{eq14}), and Eq. (\ref{Gammaf}) with  $\Delta t= 10^{-5}$. For curve \textit{a}, $\omega_s=1$, and $b=1$; for curve \textit{b}, $\omega_s=5$ and $b=1/2$.

In Fig.~\ref{fig2} we plot the diffusion coefficient $D(t)$ as a function of time $t$. The oscillatory curves corresponds to the numerical solution and  are calculated from the data of
 Fig. 1. The   curves without oscillations correspond to the analytical asymptotic limit,  Eq.~(\ref{eq9b}), with memory  Eq.~(\ref{Gamzf}).
 Here we see  that the asymptotic curves are mean values of the oscillatory ones. In this range the fit yields For curve \textit{a},  $\lambda = 0.928 \pm 0.002$, and for curve \textit{b}, $\lambda = 0.948$ $22 \pm 0.000$ $01$. We see in curve \textit{b} that the two curves collapse onto a single one. Here the transient time $\tau$ to which we refer before Eq. (\ref{eq9b}) is a decreasing function of  $b/\omega_s$.
The value of $\lambda$ approaches the exact value $1$ as the ratio $b/\omega_s$ decreases, or as time increases.  This shows the efficiency of the scaling; even before convergence is fully established, curve \textit{a}, the asymptotic curve gives us  an average value that can be used to understand the main characteristics of the process.

Consider now $\widetilde{\Gamma}(z) =az$, exactly. That means $\widetilde{R}(z) = [(1+a)z]^{-1}$
or $R(t)=[1+a]^{-1}$, and by direct integration we get $D(t)=t/(1+a)$ exactly. This is ballistic $\alpha=2$ diffusion. If we apply Eq.  (\ref{eq9b}) we obtain the same result  with $\lambda=1$. Since from the relations (\ref{Gammag}) and (\ref{lam}) the value of $\lambda$ does not depend on $\ln{(z)}$,  this result is exactly what we get from Eq. (\ref{Gamzf}). There are important differences between the $\alpha=2^{-}$ diffusion, which according to the Khinchin theorem ~\cite{Lee07,Lapas08} is ergodic, and the
$\alpha=2$ diffusion, which does violate ergodicity. This distinction was not possible before the generalization of the diffusion exponent we present here.

\section{\label{sec5} Conclusion}

In this work we generalize the concept of the diffusion exponent, and we propose a conjecture to investigate the asymptotic limits  of anomalous diffusion, through the introduction of a time scaling factor $\lambda$. We  obtain the scaling parameter exactly and we show that it is universal and depends only on the diffusion exponent.  We analyze the ballistic diffusions $\alpha=2^{-}$ and  $\alpha=2$, both analytically and numerically. The method can be useful as well to analyze large amounts of data  in stochastic processes \cite{Medino12}, and in different fields of science where is necessary to inverse a Laplace transform of the form of Eq. (\ref{eq7}).
The phenomenon of diffusion also poses challenges in the understanding of fundamental concepts in statistical physics,
 such 
general properties as the correlation function~\cite{Vainstein06},
ergodicity ~\cite{Lapas08,Siegle10,Lee07,Burov10,Magdziarz11}, and the Khinchin theorem ~\cite{Lee07,Lapas08}. 
In nonlinear  phenomena, such as growth and etching \cite{Mello01}, analytical results are rather difficult to obtain. In this way we hope that  this work may inspire research into similar asymptotic limits.

\section*{Acknowledgements}

We thank the Brazilian agencies CNPq, CAPES, FUNCAP, and FAP/DF for financial support.

\end{document}